\documentclass[runningheads]{llncs}
\usepackage[T1]{fontenc}
\usepackage{graphicx}
\usepackage{hyperref}
\usepackage{amsmath}
\usepackage{multirow}
\usepackage{multicol}
\usepackage{tikz}     
\usepackage{subcaption}
\usepackage [english]{babel}
\usepackage [autostyle, english = american]{csquotes}
\MakeOuterQuote{"}
\DeclareMathOperator*{\argmax}{argmax} 
\usepackage{algorithm}
\usepackage{algpseudocode}
\usepackage[T1]{fontenc}

\begin{document}

\title{MAPX: An explainable model-agnostic framework for the detection of false information on social media networks}

\titlerunning{MAPX: detection of false information}

\author{Sarah Condran\inst{1}\orcidID{0000-0001-7813-2116} \and
Michael Bewong\inst{1,2}\orcidID{0000-0002-5848-7451} \and
Selasi Kwashie\inst{2}\orcidID{0000-0003-4014-4976} \and
Md Zahidul Islam\inst{3}\orcidID{0000-0002-4868-4945}\and
Irfan Altas\inst{1} \and
Joshua Condran\inst{4}
}

\authorrunning{Condran et al.}
\institute{School of Computing, Mathematics and Engineering, Charles Sturt University,  Wagga Wagga, NSW, 2650, Australia \and
Artificial Intelligence and Cyber Futures Institute, Charles Sturt University, Bathurst, NSW, 2795, Australia \and
School of Computing, Mathematics and Engineering, Charles Sturt University,  Bathurst, NSW, 2795, Australia \and
Wenonah Rock Pty Ltd, Coffs Harbour, NSW, 2450, Australia 
}

\maketitle              

\begin{abstract}
The automated detection of false information has become a fundamental task in combating the spread of "fake news" on online social media networks (OSMN) as it reduces the need for manual discernment by individuals. In the literature, leveraging various content or context features of OSMN documents have been found useful. However, most of the existing detection models often utilise these features in isolation without regard to the temporal and dynamic changes oft-seen in reality, thus, limiting the robustness of the models. Furthermore, there has been little to no consideration of the impact of the quality of documents' features on the trustworthiness of the final prediction. In this paper, we introduce a novel model-agnostic framework, called MAPX, which allows evidence based aggregation of predictions from existing models in an explainable manner. Indeed, the developed aggregation method is adaptive, dynamic and considers the quality of OSMN document features. Further, we perform extensive experiments on benchmarked fake news datasets to demonstrate the effectiveness of MAPX using various real-world data quality scenarios. Our empirical results show that the proposed framework consistently outperforms all state-of-the-art models evaluated. For reproducibility, a demo of MAPX is available at \href{https://github.com/SCondran/MAPX_framework}{this link}.

\end{abstract}

\section{Introduction}

\setlength{\textfloatsep}{10pt}

The creation and spread of false information (\emph{aka} fake news) is rapidly increasing, with online social media networks (OSMN) such as Twitter, Facebook, and Weibo contributing to its rise. False information can, often unbeknownst to them, manipulate how an individual responds to topics on health, politics, and social life. 
One such example is the tweet which purported disinfectant as a cure to covid-19, leading to the mass poisoning and death of over 700 Iranians  \cite{ABC2020_corona} \cite{Farley2020_Trump}.
The term \emph{fake news} has often been used as a misnomer for documents containing false facts or data. In this work, we adopt the term \textit{false information} to preserve the generality of its application to \emph{intent}~\cite{Zhou2020_review}, \emph{outcome}~\cite{Shu2020_HPFN}, and \emph{verifiability}~\cite{Zhou2020_SAFE} of a published document. 

OSMN enables anyone to access the latest information in a variety of formats (\emph{i.e.} news articles, blogs, etc) and sources (\emph{i.e.} news outlets, public figures, etc.). 
This unregulated creation and spread of information place the onus of validating the truthfulness (or falsehood) of the information on the individual. 
However, an individual's ability to identify falsehoods objectively is influenced by factors such as \textit{confirmation bias} which makes one trust and accept information that confirms their preexisting beliefs \cite{Nickerson1998}, and \textit{selective exposure} which is when one prefers to consume information that aligns with their beliefs  \cite{Freedman1965}. External factors such as the \textit{bandwagon effect} which motivates one to perform an action because others are doing it \cite{Leibenstein1950}, and the \textit{validity effect} where one believes information after repeated exposure \cite{Boehm1994} play a critical role. 

The limitations of \emph{human-driven} false information detection is well documented. These limitations include scalability \cite{Tschiatschek2018} and bias \cite{Tchakounte2020,Wallace2022_DLA}. To overcome these limitations, AI-based decision support techniques have been proposed. These often rely on (1) \emph{content} \emph{i.e.} information contained within the document such as the words, images, emotion and publisher information \cite{Ghanem2021_FakeFlow,Shu2019_dEFEND,Wang2024_EFND,Zhang2024_HeteroSGT,Zhang2021_DEF} or (2) \emph{context} \emph{i.e.} information associated with the document such as its propagation, and credibility of its users \cite{Bing2022,Chowdhury2020_CSM,Min2022_PSIN,Ruchansky2017_CSI,Shu2019_TriFN,Silva2021,Soga2024_stance}.

Content-based approaches can be applied prior to the document's publication to pre-emptively forestall any dissemination of false information. However content-based approaches are prone to adversarial manipulation of linguistic and stylistic features of content to evade detection, and a reliance on knowledge bases for validation which may not be relevant nor available for emerging topics \cite{Huang2021_DAFD,Jin2022_FinerFact,Shu2017,Yuan2021_DAGAnn}. On the other hand, context-based approaches are often independent of linguistic and stylistic features, and knowledge bases. However, their reliance on the information generated when the document is published on an OSMN means that they are often less effective at mitigating the spread and negative consequences of false information. Further, the context information required such as propagation network may not always be available nor complete~\cite{Liu2018_PPC,Min2022_PSIN,Silva2021,Soga2024_stance,Song2021_TGFN}. Although some hybrid models such as \emph{dEFEND}~\cite{Shu2019_dEFEND}, \emph{TriFN}~\cite{Shu2019_TriFN} and \emph{CSI}~\cite{Ruchansky2017_CSI} have been proposed to combine content and context information and enable increase depth of analysis, they use a limited selection of features, making them prone to loss of reliability and effectiveness as demonstrated in Section~\ref{section_experiments}. For example, \emph{TriFN}\cite{Shu2019_TriFN} incorporated three features from both content and context to provide new perspectives on a document. The content features used were the publisher's partisan bias and the document's semantics, while the context feature was user credibility based on user similarity. The combination of features provides an increased depth of understanding. However, this feature combination overlooks critical aspects of the evolving nature of the document. Ultimately, existing detection approaches do not capture the complex and dynamic nature of false information. While hybrid models offer improvements, the reliance on a limited set of features reduces the reliability and effectiveness across various scenarios. 

In this work, we present a novel framework termed Model-agnostic Aggregation Prediction eXplanation (MAPX) which effectively derives synergy from both content and context based models. More specifically:
    (1) We design a framework MAPX which is model-agnostic and supports the integration of multiple false information detection models in parallel. 
    To the best of our knowledge, this is the first work to propose a synergistic approach to integrating false information detection models. 
    (2) We develop a novel algorithm called Dynamic Adaptive Prediction Aggregation (DAPA) which satisfies the MAPX framework and integrates base models \emph{dynamically} based on the reliability of the models. In doing so, we propose a new reliability score for assessing base models. DAPA also adopts an \emph{adaptive} approach to further moderate the contribution of each base model based on the quality of information associated with each instance of a document.
    (3) We develop a novel explainer called Hierarchical Tiered eXplanation (HTX) which, unlike existing explainable models, provides a more granular approach to improve the trustworthiness and explainability of prediction outcomes. 
    (4) We conduct extensive experiments on 3 real world datasets and a comparison with 7 state-of-the-art (SOTA) techniques to demonstrate the effectiveness, robustness and explainability of MAPX. In general, MAPX yields comparatively higher performance, which is maintained even when the quality of the prediction features deteriorates while the performance of the SOTA diminishes several fold.

\section{Related Works}
False information detection methods can be grouped into two main types: content-based and context-based. Content-based methods extract features from the content of a document such as semantics, visuals, and knowledge to train the model. For instance, \cite{Zhang2024_HeteroSGT} proposed HeteroSGT, a method using heterogeneous subgraph transformers to detect false information. This work utilises macro-level semantic information and knowledge to explore the relationship between a document, its topic, and the features associated with them. \cite{Ghanem2021_FakeFlow} proposed a method which uses a Bi-GRU network to map the flow of emotions through a document to detect false information. Another work, \cite{Rani_FNNet}, presented an ensemble-based approach named FNNet, which initialises blockchain-based deep learning to dynamically train a false information detection model. Further, \cite{Huang2024_FakeGPT} utilises large language models (LLM) in the model FakeGPT to create a reason-aware prompt method employing ChatGPT to detect false information. These content based methods cover a large range of features to pre-emptively detect false information, but are prone to adversarial manipulation with content created to mimic true information patterns to avoid detection. 
 
Context-based methods extract features from the dissemination of a document across an OSMN. For instance, \cite{Chowdhury2020_CSM} uses a statistical relational learning framework (PSL) to infer credibility scores of the user based on historical behaviours. \cite{Soga2024_stance} adapts a graph transformer network to combine the global structural information and the stance of a user's comment to enhance false information detection. To learn the user's stance BERT, an LLM, is used. Another work, \cite{Min2022_PSIN}, proposes PSIN a false information detection method which adopts a divide-and-conquer strategy to model the interactions between an item and the user for a topic-agnostic false information detection model. Context-based methods require features which due to the dynamic nature of OSMNs may not be available nor complete. 

Once a model produces a prediction, understanding how the model arrived at its decision is important to provide human moderators confidence in the decision as well as an affected user a justification. This concept termed explainability \cite{Balkır2022} has been considered in some false information detection techniques \cite{Ayoub2021,Shu2019_dEFEND,Tian2020_QSAN,Yang2022_SureFact} and it is often achieved in two main ways: intrinsic explainability and post-hoc explainability. 
Intrinsic explainability refers to self-explaining models that identify the top contributing features towards a prediction. For example, the detection framework Surefact \cite{Yang2022_SureFact} generates a heterogeneous graph of important nodes to provide insight into the model and dataset. However, this explanation has limited nodes which does not fully support human understanding. For example, the 12 node subgraph explanation for a document, which depicts connections between claim, post, keyword, and user, requires domain expertise to interrogate the prediction. While the false information detection model QSAN \cite{Tian2020_QSAN} identifies the top user comments which supports or opposes a document's claim as a means of explaining its predicted label. 
Post-hoc explainability requires a meta-model to provide an explanation of a false information detection model. Common meta-models include LIME \cite{Riberio2016_LIME} and SHAP \cite{Lundbery2017}. For example, \cite{Ayoub2021} utilised SHAP to identify the individual contributions of each word towards its final prediction of a target document's falsehood. However, these existing techniques do not offer a granular breakdown of how a model's workings contribute to its prediction for any given prediction. 

\begin{figure}[t]
    \centering
    \includegraphics[scale=0.07]{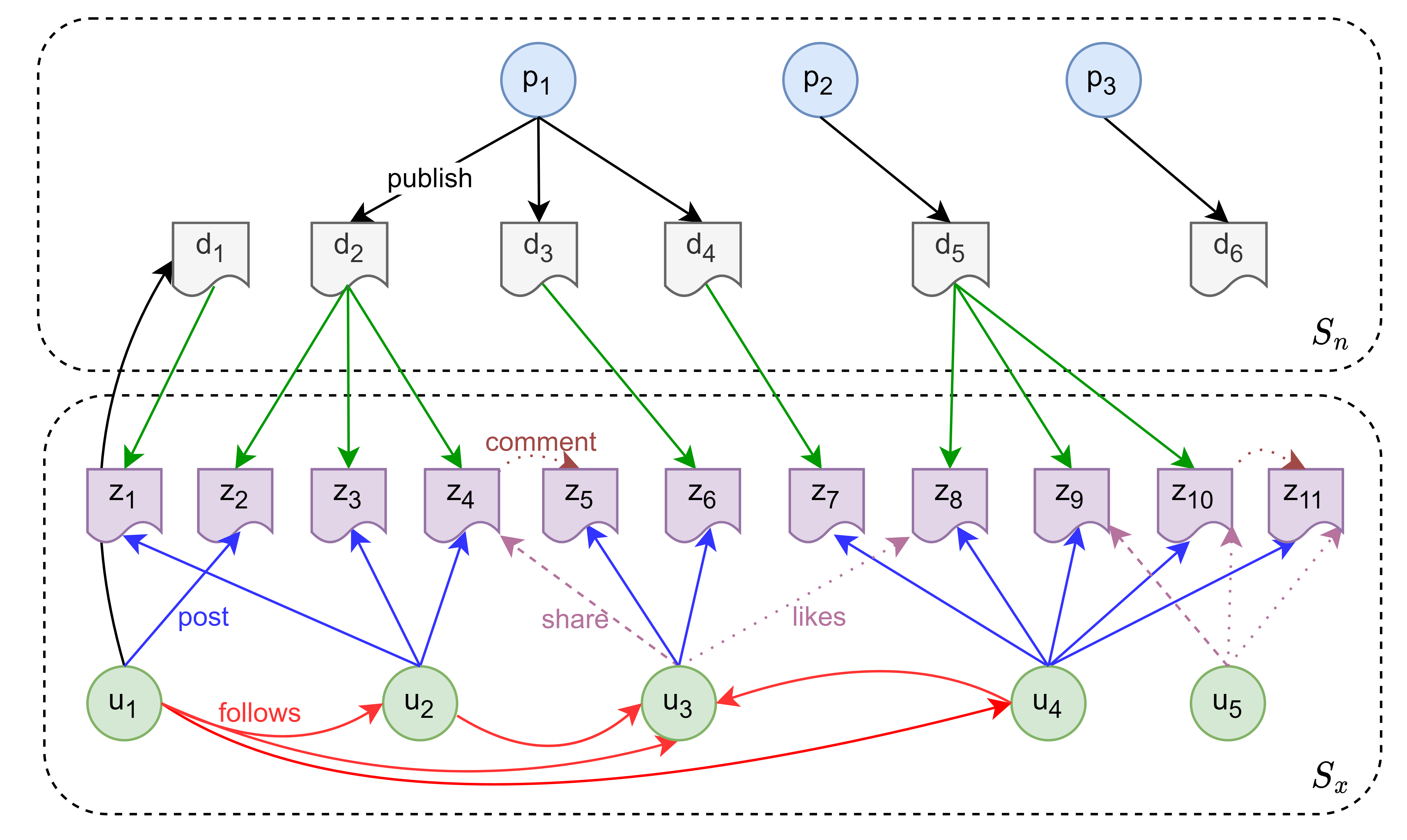}
     \vspace{-0.75em}
    \caption{Online social media network}
    \label{figureNetwork}
\end{figure}

\section{False Information Detection}\label{section_framework}
An online social network can be represented by a content network bipartite graph $\mathcal{S}_n$ and context network heterogeneous information network graph $\mathcal{S}_x$ as shown in Figure \ref{figureNetwork} and defined as follows.

\begin{definition}[Content Network]
    A content network $\mathcal{S}_n$ is defined as a bipartite graph $\mathcal{S}_n = (P,D)$, where $P$ is a set of publishers $\{p\}$, $D$ is  a set of documents $\{d\}$ and $(p,d)$ denotes a directed edge representing a publisher $p$ who publishes a document $d$.
\end{definition}

\begin{definition}[Context Network]
   A context network denoted $\mathcal{S}_x$ is defined as a heterogeneous graph $\mathcal{S}_x = (V,E,\mathcal{L}, \mathcal{F})$, where (1) $V$ is a set of nodes $\{v\}$; (2) $E$ is a set of edges $\{e\}$; (3) $\mathcal{L}(v)$ (resp. $\mathcal{L}(e)$) is a function that returns the label $l \in L$ of a node $v$ (resp. edge $e$) from the universal set of labels $L$; and (4) every node $v$, has an associated list $\mathcal{F}(v) = [(a_1,c_1) \cdots (a_n, c_n)]$ of attribute-value pairs, where $c_i \in C$ is a constant,  $a_i \in A$ is an attribute of $v$, written as $v.a_i = c_i$ , and $a_i \neq a_j$ if $i \neq j$.
\end{definition}

The labeling function $\mathcal{L}$ determines the type of node or edge. A node $v$ can be a \emph{user} or an \emph{item}  $\mathcal{I}$ where an item represents any primary or secondary content of a document created by a user. We let $\{z_1, \cdots, z_{k_i}\}_i$ denote the set of items associated with the document $d_i$. Further, an edge $e$ may represent a \emph{like}, \emph{share} or a \emph{post} between user and item nodes, \emph{comment} or \emph{share} (with comment) between \emph{item} nodes or \emph{friendship} between user nodes. 

\begin{definition}[Problem Definition]
    Let  $\mathcal{S}$ be an online social media network comprising of a content network $\mathcal{S}_n = (P,D)$ and a context network $\mathcal{S}_x = (V,E,\mathcal{L}, \mathcal{F})$, given a document $d_i \in D $ our research problem is to determine the probability $Prob_i$ of falsehood for $d_i$. 
\end{definition}

For example, considering Figure \ref{figureNetwork} we aim to calculate the probability of falsehood for $d_2$. The available information for this prediction within the content $\mathcal{S}_n$ network covers the relationship between the publisher $p_1$ and $d_2$. The context $\mathcal{S}_x$ network includes the items created on social media for each document, specifically there are three items ($z_2, z_3, z_4$) for $d_2$. There are two users ($u_1, u_2$) who create these items and two engagements on these items. In addressing the research problem there are three issues to consider. The first is how to incorporate information from both content and context networks. The second is how to identify the relevant models to analyse each instance of a document, given the available information. The third is how to provide an intrinsic explanation on the contributions of the models and information to the prediction. In the following section, we introduce our solution framework.

\section{Solution Framework} \label{section_solution}
In this section we detail our Model-agnostic Aggregated Prediction with eXplanation (MAPX) framework along with the novel DAPA aggregation algorithm and HTX explainer method. 

\subsection{Model-agnostic Aggregated Prediction with eXplanation (MAPX)} \label{subsection_mapx}
The Model-agnostic Aggregated Prediction with eXplanation (MAPX) has four main components namely (1)\emph{enricher}; (2)\emph{base modeller}; (3)\emph{aggregator}; and (4)\emph{explainer}. 

\begin{figure}[t]
    \centering
    \includegraphics[scale=0.45]{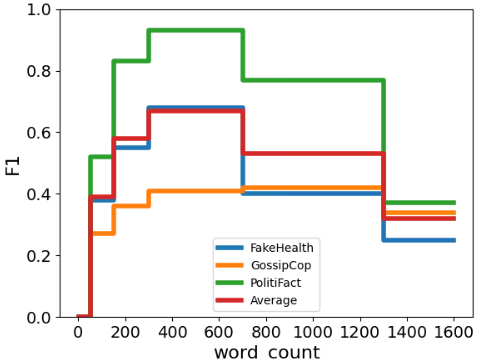}
    \vspace{-0.75em}
    \caption{Reliability factor \emph{word\_count}}
    \label{figure_F1_wordcount}
    \vspace{-2em}
\end{figure}

\begingroup
    \setlength{\tabcolsep}{15pt} 
    \begin{table}[t]
        \tiny
        \centering
        \caption{Reliability score lookup table}
        \label{table_reliability_scores}
        \begin{tabular}{|l|l|lr|}
    
            \hline
            Information & Reliability factor & \multicolumn{2}{l|}{Reliability score} \\ \hline
            \multirow{5}{*}{word} & \multirow{5}{*}{word\_count} & 0 - 25 & 0.0 \\
             &  & 26 - 100 & 0.4 \\
             &  & 101 - 300 & 0.6 \\
             &  & 301 - 600 & 0.8 \\
             &  & 601 + & 0.6 \\ \hline
            \multirow{6}{*}{publisher history} & \multirow{2}{*}{publisher\_type} & new & 0.1 \\&  & existing & 1.0 \\ \cline{2-4} 
             & \multirow{4}{*}{document\_count} & 0 - 1 & 0.1 \\
             &  & 2 - 10 & 0.4 \\
             &  & 11 - 50 & 0.5 \\
             &  & 51+ & 1.0 \\ \hline
            \multirow{12}{*}{user history} & \multirow{4}{*}{item\_count} & 0 - 1 & 0.1 \\
             &  & 2 - 10 & 0.4 \\
             &  & 11 - 50 & 0.5 \\
             &  & 51+ & 1.0 \\ \cline{2-4} 
             & \multirow{4}{*}{item\_per\_user} & 0 - 1 & 0.1 \\
             &  & 2 - 3 & 0.2 \\
             &  & 4 - 8 & 0.5 \\
             &  & 9+ & 1.0 \\ \cline{2-4} 
             & \multirow{4}{*}{document\_age} & 0 - 0.08 & 0.01 \\
             &  & 0.09 - 1 & 0.1 \\
             &  & 2 - 7 & 0.4 \\
             &  & 8+ & 1.0 \\ \hline
        \end{tabular}
    \end{table}
\endgroup

\begin{figure*}[!t]
    \centering
    \includegraphics[scale=0.0455]{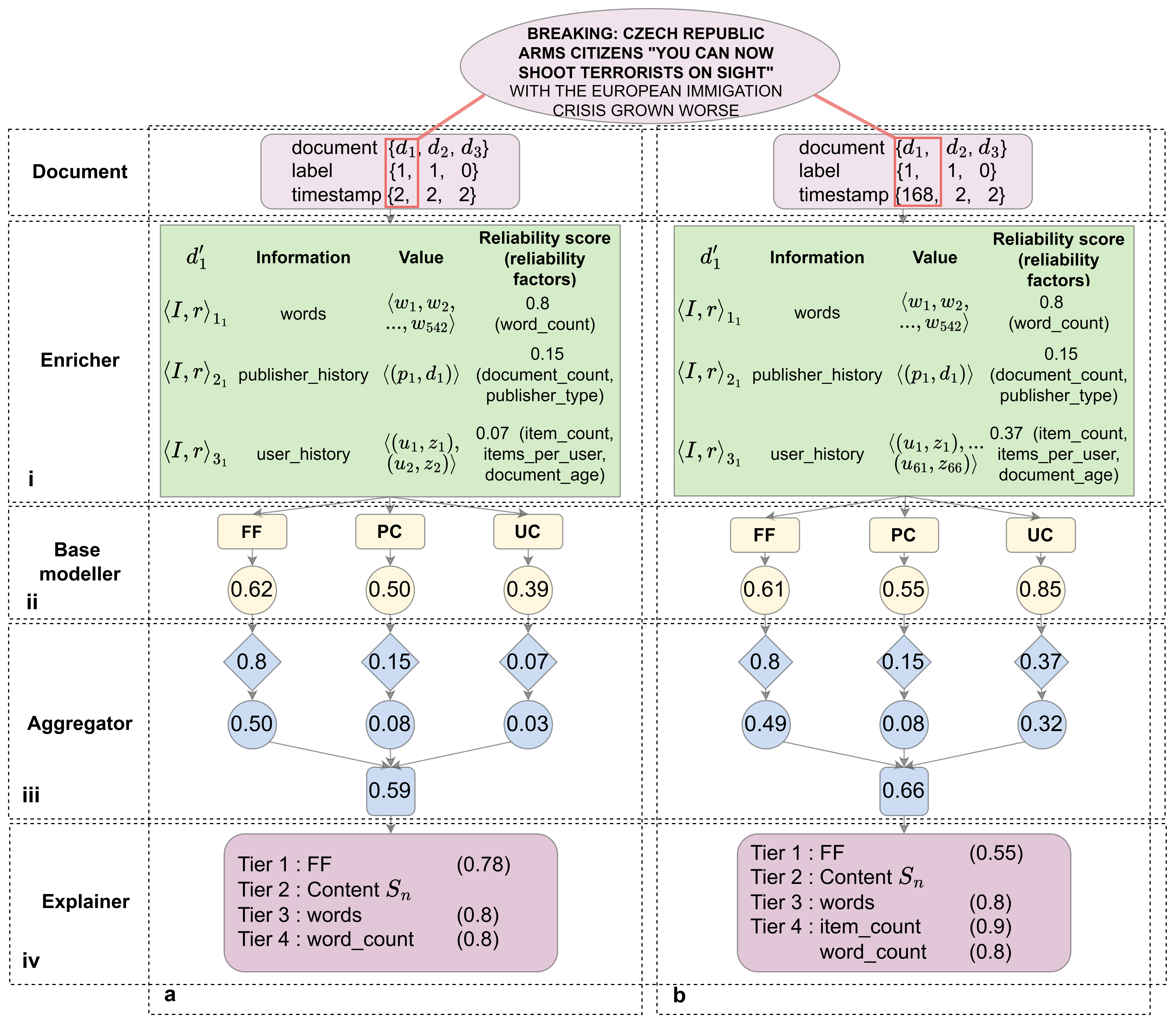}
    \vspace{-0.75em}
    \caption{The MAPX Framework}
    \label{figureFramework}
\end{figure*}

\noindent\textbf{Enricher} takes an input document $d_i$ and extracts content features, for example the historical documents $\{(p_j,d_1),\cdots (p_j,d_{m_j})\}$ of the publisher $p_j$ associated with document $d_i$; and context features for example, the set of items $\{z_1, \cdots, z_{k_i}\}_i$ associated with the document $d_i$, on an OSMN (\emph{e.g.} Twitter) to build a content $\mathcal{S}_n$ and context $\mathcal{S}_x$ network respectively. This process transforms $d_i$ into a set of information pairs $d'_i = \{\langle \mathcal{I},r\rangle_1 \cdots,  \langle\mathcal{I}, r\rangle_m\}_i$, where each pair $\langle \mathcal{I},r\rangle_j $ represents an information $\mathcal{I}$ denoting a set of features required by a base model BM and its associated reliability score $r$. 
The reliability score $r_j$ is derived from a set of reliability factors $rf$ for each information $\mathcal{I}_j$. The factors are defined based on how they influence the features within the information $\mathcal{I}_j$. For instance, for content-based models such as \emph{Fake Flow} (\emph{FF}) ~\cite{Ghanem2021_FakeFlow}, the features in ${\mathcal{I}_j}_i$ associated with $d_i$ is a set of words, thus we make the assumption that a document $d_i$ with a reasonable amount of words ${\mathcal{I}_j}_i$, provides \emph{FF} with ``sufficient'' information to make a decision on its falsehood. That is, if $|{\mathcal{I}_j}_i|$ approximates some constant $c$ then its reliability ${r_j}_i$ approximates $1$. This intuition is supported by the experiments shown in Figure \ref{figure_F1_wordcount} as measured by $F1$ scores for three datasets and different text lengths using \emph{FF}. This experiment also empirically informs the choice of the constant $c$. Table \ref{table_reliability_scores} summarises the various reliability scores which have also been similarly derived for the different extracted information. Where multiple reliability factors are used, the average value is taken. 
Figure \ref{figureFramework}.a.i illustrates the enricher which takes an input document $d_1$ (\emph{e.g.} a tweet) at some time $t_2$ and converts it to ${\langle I,r\rangle_1}_1$ by extracting the relevant information \emph{words}, and then calculates the reliability score of $0.8$ by considering the reliability factor \emph{word\_count} (\emph{i.e.} 542, not shown in the diagram). This is similarly done for $d_2$ and $d_3$. 

\noindent\textbf{Base modeller} in the training stage, takes a set of enriched data $\{d'_1,\cdots,d'_n\}$ associated with documents $\{d_1,\cdots,d_n\}$ and trains false information detection base models (BMs) which are then stored. During prediction stage, each base model in the \emph{base modeller} takes an enriched document $d'$ of the target document $d$ to be classified, and produces a probability of falsehood denoted $Prob$. Figure \ref{figureFramework}.a.ii illustrates three probability scores of 0.62, 0.50 and 0.39 generated by the three base models \emph{FF}, \emph{PC}, \emph{UC} respectively. Let $\{Prob_1,\cdots,Prob_k\}_i$ be the set of probabilities representing the falsehood produced by $k$ base models for document $d_i$.

\noindent\textbf{Aggregator} takes the falsehood probabilities $\{Prob_1,\cdots,Prob_k\}_i$ for a given document $d_i$ and generates a weighted average score. The weighted average score is determined by the formula ${\sum_{b=1}^{k}{(Avg(\{r\}_b}_i).{Prob_b}_i} ) / {\sum_{b=1}^{k}{{Avg(\{r\}_b}_i)}}$, where ${Prob_b}_i$ is the falsehood probability generated by base model BM$_b$ and ${Avg(\{r\}_b}_i)$ is the average of the set of reliability scores ${\{r\}_b}_i$ corresponding to the set of information ${\{\langle I,r\rangle\}_b}_i$ used by the base model BM$_b$ to calculate the probability of document $d_i$. Figure \ref{figureFramework}.a.iii shows the resulting weighted average falsehood probability of $0.59$. This implies that the document $d_1$ is considered to be false with a probability of 59\%. It is worth noting that $d_1$ in this case is a news article about \emph{an amended law enabling the shooting of terrorists on sight}, which is indeed false.

\noindent\textbf{Explainer} derives the various contributions from the individual base models towards the final falsehood probability score. In this work, we develop a novel explainer called hierarchical tiered explanation (HTX) which is illustrated  
in Figure \ref{figureFramework}.a.iv (HTX is discussed in the following section). In the figure, we observe that (1) the FF model contributes most to the decision making (tier 1); (2) the content network $\mathcal{S}_n$ is deemed most relevant to the decision making (tier 2); (3) the most important information is \emph{words} (tier 3); and (4) the most important reliability factor is \emph{word\_count} with a score 0.8. Such an explainer is relevant in providing platform moderators the ability to justify any censorship decision. 

\subsection{Dynamic Adaptive Prediction Aggregation (DAPA)} \label{subsectionDAPA}
DAPA is a dynamic aggregation technique which assigns different weights to each of the base models. However, unlike traditional dynamic aggregation techniques which weight the contribution of each base model differently based on the performance during the training process, DAPA adaptively adjusts the weights for each base model based on the reliability score associated with the instance of the document under consideration. Figure \ref{figureFramework}.a.iii illustrates the dynamic weighting of $0.8$, $0.15$ and $0.07$ for \emph{FF}, \emph{PC} and \emph{UC} respectively for the document $d_1$ at time $t_2$. In Figure \ref{figureFramework}.b.iii we observe that for the same document $d_1$ at time $t_{168}$ due to higher user engagement as a result of the length of exposure of the document on OSMN, the reliability of the user interactions has improved. Consequently the new weightings are $0.8$, $0.15$ and $0.37$ for \emph{FF}, \emph{PC} and \emph{UC} respectively, showing an improved contribution of model \emph{UC} towards the overall prediction. It is worth noting that with existing dynamic aggregation techniques, while different weights are assigned to different base models, the weighting stays the same across time stamps regardless of changes in reliability of the models. 

\subsection{Hierarchical Tiered eXplanation (HTX)} \label{subsectionHTX}
The proposed HTX method provides a hierarchical four-tiered explanation on how each model contributes to the final prediction and the impact of reliability factors on the models' performance. Tier 1 identifies the top contributing models in terms of reliability \emph{i.e.} $T1:= {\text{BM}_{b^*}}_i = \argmax_{BM_b \in \{\text{BM}\}}{Avg(\{r\}_b}_i)$, where ${\text{BM}_{b^*}}_i$ denotes the most reliable base model for predicting falsehood of document $d_i$; 
Tier 2 identifies the type of network (\emph{i.e.} content $\mathcal{S}_n$ or context $\mathcal{S}_x$) that contributes most to the most reliable base model ${\text{BM}_{b^*}}_i$ \emph{i.e.} $T2 := \mathcal{S}^i_* = {Avg(\{r\}_{b^*}}_i)$, where $\mathcal{S}^i_*$ is the network that contributes most to the base model ${\text{BM}_{b^*}}_i$ and ${Avg(\{r\}_{b^*}}_i)$ is the average of the reliability scores associated with ${\text{BM}_{b^*}}_i$ and $\mathcal{S}^i_*$ for the document $d_i$; 
Tier 3 then identifies the most reliable information $\mathcal{I}$ in ${\text{BM}_{b^*}}_i$ \emph{i.e.} $T3 := {\langle \mathcal{I},r\rangle ^*}_i \argmax_{\langle \mathcal{I},r\rangle \in \{{\langle \mathcal{I},r\rangle\} _{b^*}}_i} {Avg(\{r\}_{b^*}}_i)$, where ${\langle \mathcal{I},r\rangle ^*}_i$ is the most reliable information that contributes to the prediction of the falsehood of document $d_i$, $\{{\langle \mathcal{I},r\rangle\} _{b^*}}_i$ is all the information associated with the best performing base model ${\text{BM}_{b^*}}_i$ for the document $d_i$, and  ${Avg(\{r\}_{b^*}}_i)$ is the average of the reliability scores associated with ${\text{BM}_{b^*}}_i$ and $\mathcal{S}^i_*$;
and finally, Tier 4 identifies the most relevant reliability factors $rf$ and their corresponding scores \emph{i.e.} T4 := $\{ rf : rf \vdash$ ${\langle \mathcal{I},r\rangle ^*}_i\}$, where $rf$ $\vdash {\langle \mathcal{I},r\rangle ^*}_i$ denotes a reliability factor derivable from the most reliable information ${\langle \mathcal{I},r\rangle ^*}_i$ relevant to document $d_i$.
Unlike existing explainable models in false information detection models such as~\cite{Jin2022_FinerFact,Shu2019_dEFEND,Wu2020_EHIAN}, HTX provides more granular control to the user. For example, the user can contrast evidence provided by a \emph{context-based} model with that of a \emph{content-based} model, based on the contributions towards the decision. In Figure \ref{figureFramework}.a.iv we observe the explanation for the false prediction for document $d_1$. First Tier 1 identifies \emph{FF} contributed 78\% of the final aggregated prediction. Next, Tier 2 identifies that data from the content $\mathcal{S}_n$ network was the primary contributor. Tier 3 identifies the top contributing information to be \emph{words} which contributed to 80\% of the final prediction. Finally, Tier 4 identifies the top reliability factor as \emph{word\_count} with a reliability score of 0.8. In contrast Figure \ref{figureFramework}.a.i shows that \emph{publisher\_history} and \emph{user\_history} had low reliability scores, resulting in minimal contributions of models using this information.  

An algorithm that instantiates the MAPX framework including DAPA and HTX is presented in Algorithm 1\footnote{All source codes associated with MAPX and a demo is available at \href{https://github.com/SCondran/MAPX_framework}{this link}}.

\begin{algorithm}[t]
    \textsf{
    \caption{The MAPX framework}
    \textbf{Input:} A set of documents $D \in \{d_i\}$; A set \{BM\} of base models; an online social media network $\mathcal{S}$ \begin{tiny} \Comment{BM can be pre-trained or trained \emph{in situ}} \end{tiny}  \\
    \textbf{Output:} A set of probabilities $\{Prob_i\}$ representing the probability of falsehood corresponding to the set of input documents $\{d_i\}$; A set of quadruples $\{(T1,T2,T3,T4)_i\}$ representing the explainability corresponding to the set of input documents $\{d_i\}$ \\
    1. Generate content network $S_n$ and context network $S_x$ from $\mathcal{S}$ and D \\
    2. for $d_i \in D$ \\
    \hspace*{\algorithmicindent} Transform $d_i$ into $d'_i = \{\langle \mathcal{I},r\rangle_1 \cdots,\langle\mathcal{I}, r\rangle_m\}_i$ using enricher function 
        \begin{tiny} \Comment{\emph{cf.} Section \ref{subsection_mapx}} \end{tiny}  \\
    \hspace*{\algorithmicindent} With $\textbf{\text{BM}}=\{\text{BM}_1, \dots, \text{BM}_k\}$ generate the set $\{{Prob}_1, \dots, {Prob}_k\}_i$ where each \\\hspace*{\algorithmicindent}\hspace*{\algorithmicindent}   ${Prob_b}_i$ corresponds to $\text{BM}_b \in \textbf{\text{BM}}$ for $d_i$   
        \begin{tiny} \Comment{if pre-trained \textbf{\text{BM}} is not available, training \textbf{\text{BM}} using relevant $\mathcal{I}$ in \\\hspace*{\algorithmicindent}\hspace*{\algorithmicindent}\hspace*{\algorithmicindent}\hspace*{\algorithmicindent} training samples $\{d'_j\}$ and corresponding labels is required} \end{tiny} \\
    \hspace*{\algorithmicindent} $Prob_i \leftarrow DAPA ((\{{Prob}_1, \dots, {Prob}_k\}_i),d'_i)$ 
        \begin{tiny} \Comment{aggregate the set $\{{Prob}_1, \dots, {Prob}_k\}_i$ using DAPA \\\hspace*{\algorithmicindent}\hspace*{\algorithmicindent}\hspace*{\algorithmicindent}\hspace*{\algorithmicindent} function (\emph{cf.} Section \ref{subsectionDAPA})} \end{tiny} \\
    \hspace*{\algorithmicindent} Generate the quadruple (T1, T2, T3, T4)$_i$ $\leftarrow HTX ((\{{Prob}_1, \dots, {Prob}_k \}_i),d'_i)$
        \begin{tiny} \Comment{\\\hspace*{\algorithmicindent}\hspace*{\algorithmicindent}\hspace*{\algorithmicindent}\hspace*{\algorithmicindent}generate tiered explanation $\{{Prob}_1, \dots, {Prob}_k \}_i$ using the HTX function (\emph{cf.} Section \ref{subsectionHTX})} \end{tiny} \\
    3. Return $\{Prob_i\}$, \{(T1, T2, T3, T4)$_i$\}, the set of probabilities and its corresponding explanation respectively. 
    }
\end{algorithm}

\section{Empirical Evaluation} \label{section_experiments}
In this section we report on a series of experiments which validate the effectiveness of the MAPX framework. 

\noindent\textbf{Evaluation Criteria:} (1) To demonstrate the effectiveness of MAPX; (2) To assess the robustness of MAPX (with DAPA) in comparison with baseline models;  and (3) To assess the informativeness of HTX.

\noindent\textbf{Datasets:} 
In line with existing SOTA \cite{Chowdhury2020_CSM,Ghanem2021_FakeFlow,Huang2024_FakeGPT,Ruchansky2017_CSI,Shu2019_dEFEND,Shu2019_TriFN} MAPX is evaluated on the benchmark datasets \emph{PolitiFact} \cite{Shu2020_FakeHealthNetDataset}, \emph{GossipCop} \cite{Shu2020_FakeHealthNetDataset}, and \emph{FakeHealth} \cite{Dai2020_FakeHealthDataset} which relate to politics, social life and health respectively (\emph{c.f.}~Table \ref{tableDataset} for details).

\begingroup
    \setlength{\tabcolsep}{10pt} 
    \begin{table}[tbp]
        \vspace{-1.5em}
        \tiny
        \begin{center}
        \caption{Datasets}
        \begin{tabular}{|l|l|l|l|}
            \hline                               & \textbf{PolitiFact} & \textbf{GossipCop} & \textbf{FakeHealth} \\ \hline
            \textbf{Document count  }                 & $642$        &  $20645$    & $1626 $      \\
            \textbf{True : Fake split}                      & $46:54 $     &  $77:23$    & $72:28$      \\
            \textbf{Total \# publishers }                 & $366$        & $2009 $     & $85$         \\
            \textbf{Total \# items}                       & $511,044$    & $1,458,842$  & $1,401,886 $   \\
            \textbf{Total \# users}                       & $292,790$    & $313,878$  & $313,916$    \\
            \hline
        \end{tabular}
        \label{tableDataset} 
        \end{center}
    \end{table}
\endgroup

\begingroup
    \setlength{\tabcolsep}{2.2pt} 
    \begin{table}[t]
        \tiny
        \centering
        \caption{Effectiveness of MAPX}
        \label{table_results_model}
        \begin{tabular}{|l|l|lllll|}

            \hline
            \multicolumn{1}{|c|}{} &  & \multicolumn{1}{c}{\textbf{MAPX}} & \textbf{dEFEND} \cite{Shu2019_dEFEND} & \textbf{CSI} \cite{Ruchansky2017_CSI}\cite{Shu2019_dEFEND} & \textbf{TriFN} \cite{Shu2019_TriFN} & \textbf{FakeGPT} \cite{Huang2024_FakeGPT} \\ \hline
            \multirow{2}{*}{\textbf{PolitiFact}} & Acc & $\textbf{0.93}$ & $0.90$ & $0.83 $& $0.86$ & $0.69$\\
             & F1 & $\textbf{0.93}$ & $\textbf{0.93}$ & $0.87$ & $0.87$ & $0.66$\\ \hline
            \multirow{2}{*}{\textbf{GossipCop}} & Acc & $\textbf{0.92}$  & $0.81$ & $0.77 $& NA & $0.69$\\
             & F1 & $\textbf{0.83}$ & $0.76$ & $0.68 $& NA & $0.66$\\ \hline 
        \end{tabular}
    \end{table}
\endgroup

\noindent\textbf{Baseline Models:} In total seven state-of-the-art baseline models are considered: (1) \emph{Fakeflow} (FF) \cite{Ghanem2021_FakeFlow} utilises content features with Bidirectional Gated Recurrent Units (Bi-GRUs) to learn the flow of affective information throughout a document; (2) \emph{Publisher Credibility} (PC) adapted from \cite{Chowdhury2020_CSM} utilises document publisher features to train a probabilistic soft logic (PSL) model to calculate the credibility of a publisher; (3) \emph{User Credibility} (UC) adapted from \cite{Chowdhury2020_CSM} utilises context features in the form of users who create an item to train a PSL model to calculate the credibility of that user; (4) \emph{dEFEND} \cite{Shu2019_dEFEND} combines the document's content with user comments to lean co-attention networks that connect user comments to claims within a document; (5) \emph{CSI} \cite{Ruchansky2017_CSI} utilises both content and context features to first train a recurrent neural network to learn the temporal user activity patters, then learn a user suspicious score; (6) \emph{TriFN} \cite{Shu2019_TriFN} combines content features, in the form of publisher partisan bias and document semantics, along with context features, in the form of user credibility, to create embeddings to train a semi-supervised linear classifier; and (7) \emph{FakeGPT} \cite{Huang2024_FakeGPT} utilises content features with LLMs to create reason aware prompts to analyse a document.  In our experiments, we purposefully select \emph{FF}, \emph{PC}, \emph{UC} as representative base models for MAPX.

All experiments were conducted on Ubuntu Linux using Cisco UCSC-C220-M5SX servers with 80 cores and 192GB ram. The framework was built using python 3 running on Linux. For all BMs the train-validation-test split is 70-10-20, with 10-fold cross validation. 

\noindent\emph{\textbf{Effectiveness of MAPX}.} In this experiment, we assess the effectiveness of MAPX (with DAPA) in comparison with a large language model based method \emph{FakeGPT}~\cite{Huang2024_FakeGPT}, and three SOTA hybrid models \emph{dEFEND}~\cite{Shu2019_dEFEND}, \emph{CSI}~\cite{Ruchansky2017_CSI}, and \emph{TriFN}~\cite{Shu2019_TriFN}. Both \emph{TriFN} and \emph{CSI} incorporate \emph{words}, \emph{publisher\_history}, and \emph{user\_history} information while \emph{dEFEND} incorporates \emph{words} and \emph{user\_history}. Table ~\ref{table_results_model} presents the accuracy and F1 scores for MAPX, and the reported results of the \emph{FakeGPT}, \emph{dEFEND}, \emph{CSI}, and \emph{TriFN} for the \emph{PolitiFact} and \emph{GossipCop} datasets. The results show that MAPX significantly outperforms all the baseline techniques.

We point out that, a time efficiency analysis has not been included in our evaluation since the computational complexity of MAPX is equivalent to the worst performing base model in MAPX.

\begin{table}[!t]
    \vspace{-2em}
    \tiny
    \centering
    \caption{Impact of feature reliability on F1}
    \label{table_results}
    \begin{tabular}{|l|l|l|l|}
        \hline
        & \textbf{PolitiFact} & \textbf{GossipCop} & \textbf{FakeHealth} \\ \hline
        \multicolumn{4}{|c|}{\textbf{a. publisher\_type}} \\ \hline
        \textbf{MAPX - DAPA} & \tikz[baseline]{\node[draw=red, thick, rounded corners, shape=rectangle] {$ 0.94\xrightarrow{}0.93$};} $(0.01) [2]$ & \tikz[baseline]{\node[draw=red, thick, rounded corners, shape=rectangle] {$ 0.84\xrightarrow{}0.83$};}$(0.01) \textbf{[1]}$ & \tikz[baseline]{\node[draw=red, thick, rounded corners, shape=rectangle] {$ 0.28\xrightarrow{}0.67$};}$(0) \textbf{[1]}$ \\ \hline
        \textbf{MAPX - BMAcc} & $ 0.92\xrightarrow{}0.91(0.01) \textbf{[2]}$ & $ 0.72\xrightarrow{}0.71(0.01) \textbf{[1]}$ & $ 0.25\xrightarrow{}0.68(0) \textbf{[1]}$ \\ \hline
        \textbf{MAPX - Max} & $ 0.92\xrightarrow{}0.91(0.01) \textbf{[2]}$ & $ 0.72\xrightarrow{}0.66(0.06) [4]$ & $ 0.10\xrightarrow{}0.53(0) \textbf{[1]}$ \\ \hline
        \textbf{MAPX - Av} & $ 0.93\xrightarrow{}0.92(0.02) [3]$ & $ 0.70\xrightarrow{}0.66(0.04) [3]$ & $ 0.05\xrightarrow{}0.41(0) \textbf{[1]}$ \\ \hline
        \textbf{FF} & $ 0.60\xrightarrow{}0.85(0) \textbf{[1]}$ & $ 0.56\xrightarrow{}0.42(0.14) [5]$ & $ 0\xrightarrow{}0(0) \textbf{[1]}$ \\ \hline
        \textbf{PC} & $ 0.85\xrightarrow{}0.69(0.16) [4]$ & $ 0.57\xrightarrow{}0.40(0.17) [6]$ & $ 0.25\xrightarrow{}0.60(0) \textbf{[1]}$ \\ \hline
        \textbf{UC} & $ 0.85\xrightarrow{}0.89(0) \textbf{[1]}$ & $ 0.84\xrightarrow{}0.82(0.02) [2]$ & $ 0.05\xrightarrow{}0.03(0.02) [2]$ \\ \hline
    
        \multicolumn{4}{|c|}{\textbf{b. items\_per\_user}} \\ \hline
        \textbf{MAPX - DAPA} & \tikz[baseline]{\node[draw=red, thick, rounded corners, shape=rectangle] {$ 0.93\xrightarrow{}0.93$};}$(0) \textbf{[1]}$ & \tikz[baseline]{\node[draw=red, thick, rounded corners, shape=rectangle] {$ 0.84\xrightarrow{}0.78$};}$(0.06) [2]$ & \tikz[baseline]{\node[draw=red, thick, rounded corners, shape=rectangle] {$ 0.32\xrightarrow{}0.32$};}$(0) \textbf{[1]}$ \\ \hline
        \textbf{MAPX - BMAcc} & $ 0.92\xrightarrow{}0.92(0) \textbf{[1]}$ & $ 0.72\xrightarrow{}0.61(0.11) [5]$ & $ 0.29\xrightarrow{}0.28(0.01) [2]$ \\ \hline
        \textbf{MAPX - Max} & \tikz[baseline]{\node[draw=red, thick, rounded corners, shape=rectangle] {$ 0.93\xrightarrow{}0.93$};}$(0) \textbf{[1]}$ & $ 0.71\xrightarrow{}0.60(0.11) [5]$ & $ 0.13\xrightarrow{}0.13(0) \textbf{[1]}$ \\ \hline
        \textbf{MAPX - Av} & \tikz[baseline]{\node[draw=red, thick, rounded corners, shape=rectangle] {$ 0.93\xrightarrow{}0.93$};}$(0) \textbf{[1]}$ & $ 0.69\xrightarrow{}0.58(0.11) [5]$ & $ 0.07\xrightarrow{}0.07(0) \textbf{[1]}$ \\ \hline
        \textbf{FF} & $ 0.74\xrightarrow{}0.74(0) \textbf{[1]}$ & $ 0.54\xrightarrow{}0.48(0.06) [4]$ & $ 0\xrightarrow{}0(0) \textbf{[1]}$ \\ \hline
        \textbf{PC} & $ 0.76\xrightarrow{}0.76(0) \textbf{[1]}$ & $ 0.54\xrightarrow{}0.50(0.04) \textbf{[1]}$ & $ 0.31\xrightarrow{}0.31(0) \textbf{[1]}$ \\ \hline
        \textbf{UC} & $ 0.89\xrightarrow{}0.88(0.01) [2]$ & $ 0.84\xrightarrow{}0.77(0.07) [3]$ & $ 0.05\xrightarrow{}0.05(0) \textbf{[1]}$ \\ \hline
    
        \multicolumn{4}{|c|}{\textbf{c. items\_count}} \\ \hline
        \textbf{MAPX - DAPA} & \tikz[baseline]{\node[draw=red, thick, rounded corners, shape=rectangle] {$ 0.94\xrightarrow{}0.89$};} $(0.05) \textbf{[1]}$ & \tikz[baseline]{\node[draw=red, thick, rounded corners, shape=rectangle] {$ 0.84\xrightarrow{}0.84$};} $(0) \textbf{[1]}$ & \tikz[baseline]{\node[draw=red, thick, rounded corners, shape=rectangle] {$ 0.29\xrightarrow{}0.46$};} $(0) \textbf{[1]}$ \\ \hline
        \textbf{MAPX - BMAcc} & $ 0.93\xrightarrow{}0.87(0.06) [2]$ & $ 0.72\xrightarrow{}0.78(0.06) [4]$ & $ 0.25\xrightarrow{}0.42(0) \textbf{[1]}$ \\ \hline
        \textbf{MAPX - Max} & $ 0.93\xrightarrow{}0.88(0.05) \textbf{[1]}$ & $ 0.70\xrightarrow{}0.78(0.08) [5]$ & $ 0.11\xrightarrow{}0.22(0) \textbf{[1]}$ \\ \hline
        \textbf{MAPX - Av} & \tikz[baseline]{\node[draw=red, thick, rounded corners, shape=rectangle] {$ 0.94\xrightarrow{}0.89$};} $(0.05) \textbf{[1]}$ & $ 0.69\xrightarrow{}0.74(0.05) [3]$ & $ 0.06\xrightarrow{}0.14(0) \textbf{[1]}$ \\ \hline
        \textbf{FF} & $ 0.77\xrightarrow{}0.67(0.10) [5]$ & $ 0.54\xrightarrow{}0.57(0.03) [2]$ & $ 0\xrightarrow{}0(0) \textbf{[1]}$ \\ \hline
        \textbf{PC} & $ 0.77\xrightarrow{}0.72(0.05) [3]$ & $ 0.54\xrightarrow{}0.59(0.05) [3]$ & $ 0.28\xrightarrow{}0.46(0) \textbf{[1]}$ \\ \hline
        \textbf{UC} & $ 0.90\xrightarrow{}0.83(0.07) [4]$ & $ 0.83\xrightarrow{}0.83(0) \textbf{[1]}$ & $ 0.04\xrightarrow{}0.12(0) \textbf{[1]}$ \\ \hline
    \end{tabular}
\end{table}

\begin{figure}[!t]
  \centering  
  \begin{subfigure}[b]{0.49\textwidth}
    \centering
    \includegraphics[width=\textwidth]{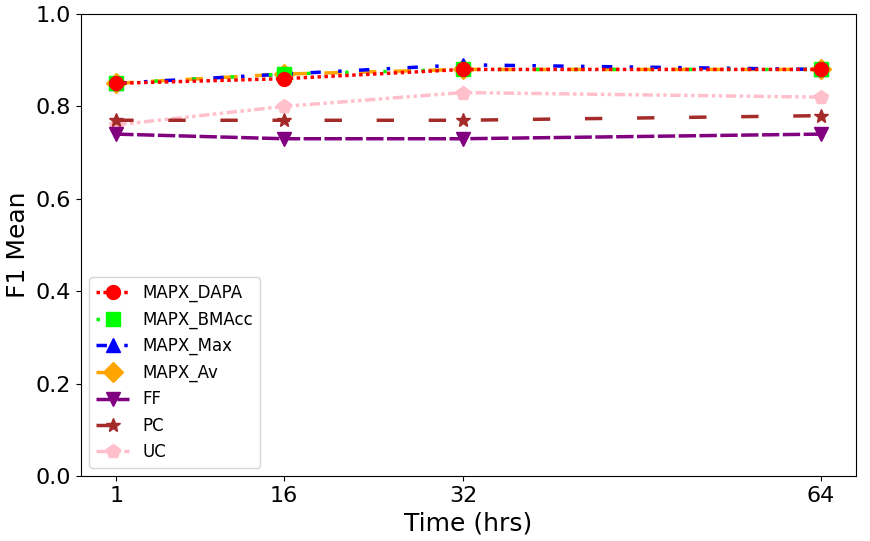}
    \caption{PolitiFact}
    \label{figureAgga}
  \end{subfigure}
  \begin{subfigure}[b]{0.49\textwidth}
    \centering
    \includegraphics[width=\textwidth]{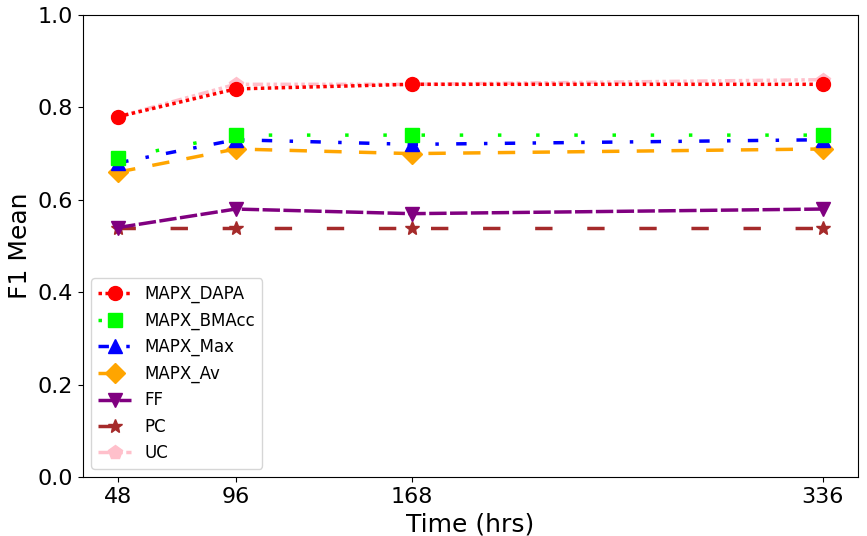}
    \caption{GossipCop}
    \label{figureAggb}
  \end{subfigure}
    \begin{subfigure}[c]{0.49\textwidth}
        \centering
        \includegraphics[width=\textwidth]{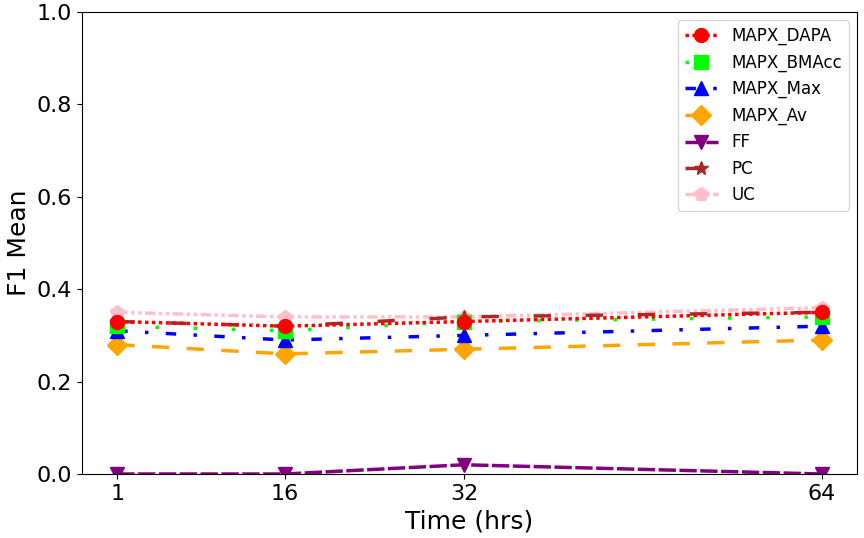}
        \caption{FakeHealth}
        \label{figureAggc}
    \end{subfigure}
    \vspace{-0.75em}
  \caption{Temporal Analysis}
  \label{figure_temporal_analysis}
\end{figure}

\noindent\emph{\textbf{Robustness of MAPX.}}
In this experiment, we assess the robustness of MAPX \emph{w.r.t.} changes in the quality of information associated with target documents. To demonstrate the impact of DAPA, we also consider various well-known aggregation functions such as BMAcc (\emph{i.e.} the base models are dynamically weighted based on their training performance), Max (\emph{i.e.} returns the probability closest to 0 or 1 as the prediction), and Av (\emph{i.e.} the average of all base model probabilities) in lieu of DAPA. We also consider the baseline models \emph{FF}, \emph{PC}, \emph{UC}. Table~\ref{table_results} shows the impact of feature reliability on various detection models.

Specifically, Table~\ref{table_results}.a represents the results of controlling the reliability factor \emph{publisher\_type}. That is, if a publisher is new (unobserved within the training data) then its reliability as a prediction feature is low, and high otherwise. For example, a model that relies on publisher credibility will be more reliable if there is sufficient historical information about the publisher of a document, however for a new publisher, its publisher features is likely to be unreliable due to lack of historical information. Similar explanations apply to \emph{items\_per\_user} and \emph{items\_count} for Table~\ref{table_results}.b and Table~\ref{table_results}.c respectively (\emph{c.f.} Table~\ref{table_reliability_scores}). Each result in the table has the format $f1 \rightarrow f1' ($\emph{diff}$)[\text{rank}]$ where $f1$ is the F1 score for the corresponding model and dataset containing only reliable features \emph{w.r.t} the reliability factor (\emph{e.g.} \emph{publisher\_type}, \emph{items\_per\_user} or \emph{items\_count} ) while $f1'$ is the F1 score when the data contains unreliable features; $($\emph{diff}$)$ is the difference between $f1$ and $f1'$; and $[\text{rank}]$ is the ranking of the corresponding model \emph{w.r.t.} other models based on $($\emph{diff}$)$. Ideally, both $f1$ and $f1'$ should be large while $($\emph{diff}$)$ is small. In Table~\ref{table_results}, we observe that MAPX, and in particular DAPA gives the best results. Implying that MAPX-DAPA is robust against changes in the reliability of the datasets. Similar results can be observed when the temporal impact on the reliability of features is taken into consideration (\emph{c.f.} Figure~\ref{figure_temporal_analysis}). That is, when a document is published on an OSMN, users' interactions and the reliability of associated features are likely to grow with time. It is worth noting that, in Figure~\ref{figure_temporal_analysis}.a the base models \emph{FF} and \emph{PC} remain constant since the reliability of their features are time-independent (\emph{i.e.} \emph{word\_count}, \emph{document\_count}, \emph{publisher\_type}), while \emph{UC}, which depends on \emph{items\_per\_user}, performance improves over time as more information becomes available. It is clear that MAPX accounts for the variation of performance due to reliability (or lack thereof) of features and provides a consistent performance over time. 

\begin{figure}[t]
    \centering
    \includegraphics[scale=0.039]{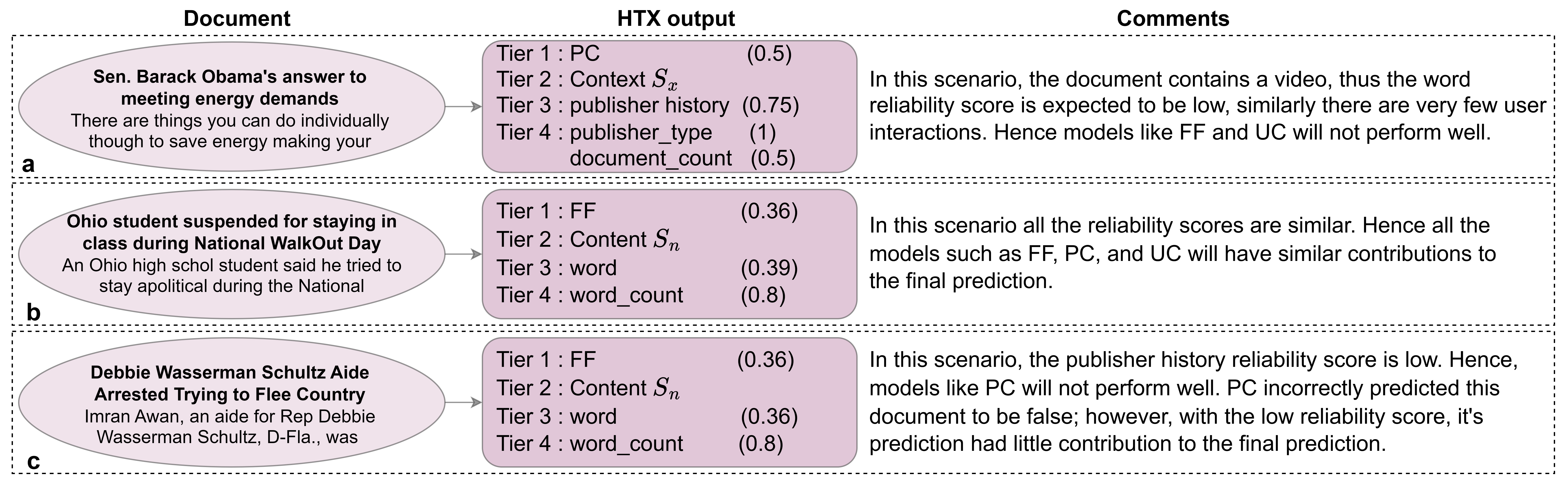}
    \vspace{-0.75em}
    \caption{HTX explanation}
    \label{figureHTX}
\end{figure}

\noindent\emph{\textbf{Qualitative Assessment of HTX.}}
In this section we show the explanations generated by HTX for three use cases in Figure \ref{figureHTX}. Consider Figure \ref{figureHTX}.a, relating to \emph{Sen. Barack Obama's answer to meeting energy demands} with an explanation for the prediction 0.29 probability of falsehood. (Note that this has been factchecked to be true). Tier 1 identified \emph{PC} as the highest contributing model, and Tier 2 identified the context $\mathcal{S}_x$ as the highest contributing network. Tier 3 identified the top contributing information to be \emph{publisher history} with a reliability score of 0.75. Tier 4 identified the most relevant reliability factors were \emph{publisher\_type} and \emph{document\_count}. That is, these factors had sufficient information available for the model to produce a reliable prediction of falsehood. Similar analysis can also be inferred from Figure \ref{figureHTX}b and c.

\section{Conclusion}
In this work, we designed a novel model-agnostic framework MAPX which improves the detection of false information by leveraging multiple base models in a dynamic fashion. Further, we proposed a novel dynamic adaptive prediction aggregation (DAPA) and a hierarchical tiered explanation (HTX) with several advantages over existing aggregators and explainers respectively, as demonstrated in the empirical evaluation. The versatility of MAPX is expected to provide platform moderators a plug-and-play solution, allowing for more effective moderation of information on OSMNs while still providing a means for moderators to interrogate the system. In future work, we plan to extend our experiments to investigate the versatility of MAPX in detecting false information across different OSMNs. Further, we plan to explore the suitability of integrating LLM models into the MAPX framework.

\bibliographystyle{splncs04}
\bibliography{paper}

\end{document}